\documentclass[journal]{IEEEtran}

\usepackage[T1]{fontenc}
\usepackage{amssymb,latexsym,graphicx,times,amsmath,subfigure,threeparttable,booktabs,bm,color,multirow,cite,amsmath,amsfonts,amsthm,pifont}
\usepackage{color}
\usepackage[lined,ruled,commentsnumbered]{algorithm2e}

\makeatletter

\newcommand{\Rmnum}[1]{\expandafter\@slowromancap\romannumeral #1@}
\makeatother

\begin{document}
\title{AFPM: Alignment-based Frame Patch Modeling for Cross-Dataset EEG Decoding}

\author{Xiaoqing~Chen, Siyang~Li and Dongrui~Wu,~\IEEEmembership{Fellow,~IEEE}

\thanks{X.~Chen, S. Li and D.~Wu are with the Key Laboratory of the Ministry of Education for Image Processing and Intelligent Control, School of Artificial Intelligence and Automation, Huazhong University of Science and Technology, Wuhan 430074, China. X. Chen and D. Wu are also with Zhongguancun Academy, Beijing, 100094 China. Email: \{xqchen914, syoungli, drwu\}@hust.edu.cn. }
\thanks{This research was supported by Zhongguancun Academy 20240301.}}

\markboth{}
{Chen \MakeLowercase{\textit{et al.}}: Alignment based Frame Patch Modeling for Cross-dataset EEG Decoding}
\maketitle

\begin{abstract}
Electroencephalogram (EEG) decoding models for brain-computer interfaces (BCIs) struggle with cross-dataset learning and generalization due to channel layout inconsistencies, non-stationary signal distributions, and limited neurophysiological prior integration. To address these issues, we propose a plug-and-play Alignment-Based Frame-Patch Modeling (AFPM) framework, which has two main components: 1) Spatial Alignment, which selects task-relevant channels based on brain-region priors, aligns EEG distributions across domains, and remaps the selected channels to a unified layout; and, 2) Frame-Patch Encoding, which models multi-dataset signals into unified spatiotemporal patches for EEG decoding. Compared to 17 state-of-the-art approaches that need dataset-specific tuning, the proposed calibration-free AFPM achieves performance gains of up to 4.40\% on motor imagery and 3.58\% on event-related potential tasks. To our knowledge, this is the first calibration-free cross-dataset EEG decoding framework, substantially enhancing the practicalness of BCIs in real-world applications.
\end{abstract}


\IEEEpeerreviewmaketitle

\section{Introduction}
\IEEEPARstart{A} brain-computer interface (BCI) enables direct communication between the human brain and external devices, supporting assistive, augmentative, and rehabilitative applications for cognitive and sensorimotor functions\cite{Ienca2018}. BCIs have demonstrated broad utility in neurological rehabilitation~\cite{Daly2008}, tactile exploration~\cite{ODoherty2011}, robotic control~\cite{Hochberg2012}, speech synthesis~\cite{Metzger2023}, and so on. Based on signal acquisition approaches, BCIs are generally classified as non-invasive, partially invasive, and invasive ones. Non-invasive BCIs, which commonly use scalp electroencephalogram (EEG) signals~\cite{NicolasAlonso2012}, are the most prevalent due to their ease of use and safety.

Recent advances in machine learning and deep learning have led to significant improvements in EEG decoding. Traditional machine learning methods, such as common spatial patterns (CSP)~\cite{Blankertz2008} and xDAWN~\cite{Rivet*2009}, extract expert-designed features from EEG data, achieving high accuracy with low computational cost. More recently, deep neural networks, including convolutional neural networks~\cite{Lawhern2018}, long short-term memory networks, graph neural networks, and Transformers~\cite{Song2023}, have further enhanced the decoding performance. To reduce the reliance on labelled target-domain EEG data, transfer learning has also been extensively explored~\cite{drwuMITLBCI2022}. However, existing transfer learning methods are often limited to one or two source datasets due to heterogeneous configurations across EEG corpora~\cite{10704990}. Moreover, they typically require calibration data from the target domain to account for inter-subject, inter-session, and/or inter-device variations~\cite{drwuMITLBCI2022}. Although these models can improve the target domain EEG decoding performance, they lack scalability and remain limited to single-domain applications.

Foundation models recently have achieved remarkable success in natural language processing, computer vision, and speech recognition, where large-scale, high-quality training data play a crucial role in model learning~\cite{Brown2020}. However, developing foundation models for EEG presents unique challenges: 1) EEG data collection is labor-intensive, resulting in small dataset sizes; 2) channel layouts and recording durations vary significantly across datasets, complicating unified representation learning; and, 3) scalp EEG exhibits low signal-to-noise ratios and substantial distributional shifts due to artifacts and variability across subjects, sessions, and recording systems.

Emerging research has begun exploring EEG foundation models for large-scale cross-dataset learning~\cite{Kostas2021,Yang2023, Jiang2024, Wang2024, Wang2025}. These studies typically pre-train models via self-supervised learning on multiple EEG datasets and then apply them to diverse downstream tasks. A common approach segments EEG signals into channel-wise temporal patches and tokenizes them to handle heterogeneous data structures. Initial results indicate that cross-dataset EEG pretraining is feasible and promising. Nonetheless, key limitations remain:
\begin{enumerate}
\item \textbf{Inadequate handling of data non-stationarity}: Existing pipelines use very basic preprocessing (e.g., filtering), which are inadequate in mitigating EEG distribution shifts across subjects, sessions, and devices, leading to sub-optimal model performance.
\item \textbf{Under-utilization of neurophysiological expertise}: Task-related activation patterns, such as event-related desynchronization/synchronization (ERD/ERS) in motor imagery~\cite{Pfurtscheller1997}, are rarely incorporated into feature encoding, leading to suboptimal representations.
\item \textbf{High calibration overhead}: Many frameworks require substantial calibration data from the target dataset, limiting their practicalness.
\end{enumerate}

To address these challenges, we propose Alignment-based Frame-Patch Modeling (AFPM), a calibration-free framework for cross-dataset EEG learning. AFPM consists of two main components:
\begin{enumerate}
\item \textbf{Spatial Alignment (SA)}: Guided by task-specific brain region priors, SA selects neurophysiologically relevant channels while filtering out redundant inputs. It then applies Euclidean alignment to harmonize the signal distributions across datasets, subjects, and sessions, followed by remapping selected channels to a unified spatial layout.
\item \textbf{Frame-Patch Encoding (FPE)}: FPE constructs synchronized spatiotemporal patches from all task-relevant channels. This approach explicitly captures coordinated neural activation patterns essential for robust decoding.
\end{enumerate}

We conduct comprehensive experiments on motor imagery (MI) and event-related potential (ERP) tasks. AFPM is pretrained on eight MI and six ERP datasets and evaluated against seventeen state-of-the-art models, including both foundation and non-foundation approaches. Without target-domain fine-tuning, AFPM achieves performance gains of up to 4.40\% on MI and 3.58\% on ERP tasks. Our key contributions are as follows:
\begin{enumerate}
\item To the best of our knowledge, AFPM is the first EEG decoding framework to achieve state-of-the-art cross-dataset performance without requiring target-domain calibration, offering practical value for scalable EEG pretraining and real-world deployment in applications such as stroke rehabilitation.
\item We introduce SA, a neurophysiology-guided alignment mechanism that enhances cross-domain generalization by selecting and mapping informative channels and mitigating low signal-to-noise challenges.
\item We propose FPE, an encoding paradigm that captures critical spatiotemporal interactions, improving both decoding accuracy and computational efficiency.
\end{enumerate}

The remainder of this paper is organized as follows: Section~\ref{sect:rw} reviews related work. Section~\ref{sect:atad} details the AFPM framework. Section~\ref{sect:es} presents experimental results. Section~\ref{sect:CFR} concludes the paper.

\section{Related Work} \label{sect:rw}
This section reviews prior work on EEG decoding, transfer learning and EEG foundation models.

\subsection{EEG Decoding}
EEG decoding methods can be broadly categorized into three groups:
\begin{enumerate}
\item Traditional feature engineering. CSP~\cite{Blankertz2008} maximizes inter-class variance to enhance motor imagery discrimination, while xDAWN~\cite{Rivet*2009} improves the signal-to-noise ratio for ERP detection.
\item Deep learning models. EEGNet~\cite{Lawhern2018} uses depthwise separable convolutions tailored for EEG data. DeepCNN and ShallowCNN~\cite{Schirrmeister2017} incorporate convolutional blocks optimized for EEG signals. SPaRCNet~\cite{Jing2023} and ContraWR~\cite{Yang2023a} leverage residual connections within convolutional structures. IFNet~\cite{Wang2023d} and FBCNet~\cite{Mane2021} apply fixed-band filters to enable multi-band signal decomposition, treating filtered outputs as multi-view representations.
\item Transformer-based Architectures. ST-Transformer~\cite{Song2021} performs spatiotemporal fusion through hierarchical attention. EEGConformer~\cite{Song2023} integrates convolutional layers with self-attention mechanisms, while EEGDeformer~\cite{Ding2025} adopts a coarse-to-fine transformer framework enhanced with an information purification module.
\end{enumerate}

\subsection{Transfer Learning}
Transfer learning techniques have been extensively explored to address cross-domain variability in EEG decoding. These methods are generally classified into two categories:
\begin{enumerate}
\item Homogeneous transfer learning, which targets inter-session and inter-subject distributional shifts. Approaches in this category include input-space alignment methods such as Euclidean alignment~\cite{He2019}, which normalize covariance matrices across subjects. Additionally, feature-space adaptation methods aim to align the marginal and conditional distributions between source and target domains to improve generalization~\cite{Zhuang2021}.
\item Heterogeneous transfer learning, which addresses structural discrepancies across devices. These methods typically require target-domain data to map both training and testing distributions into a unified representation space~\cite{10704990}. However, the heterogeneity within the training set itself when multiple diverse datasets are involved generally lacks consideration, making such approaches unsuitable for foundation model pretraining.
\end{enumerate}

\subsection{EEG Foundation Models}
EEG foundation models are pretrained on large-scale EEG corpora and adapted to diverse downstream tasks.

BENDR~\cite{Kostas2021} employs masked autoencoding combined with contrastive learning to pretrain on extensive EEG datasets. BIOT~\cite{Yang2023} discretizes each EEG channel into fixed-length segments containing localized signal patterns, enabling tokenization across varying channel counts and temporal lengths for Transformer-based decoding. LaBraM~\cite{Jiang2024} applies vector-quantized neural spectrum prediction to develop a neural tokenizer adaptable to heterogeneous EEG datasets. EEGPT~\cite{Wang2024} achieves competitive performance across three downstream tasks using multimodal self-supervised pretraining with contrastive and reconstruction losses. CBraMod~\cite{Wang2025} incorporates asymmetric conditional positional encoding into a criss-cross Transformer architecture, demonstrating state-of-the-art performance on ten downstream datasets. Brant~\cite{Zhang2023}, designed for intracranial EEG analysis, captures temporal-spectral features and exhibits strong generalization across diverse downstream tasks.

\section{Methodology} \label{sect:atad}

\subsection{Overview}
\begin{figure*}
    \centering
    \includegraphics[width=0.7\linewidth]{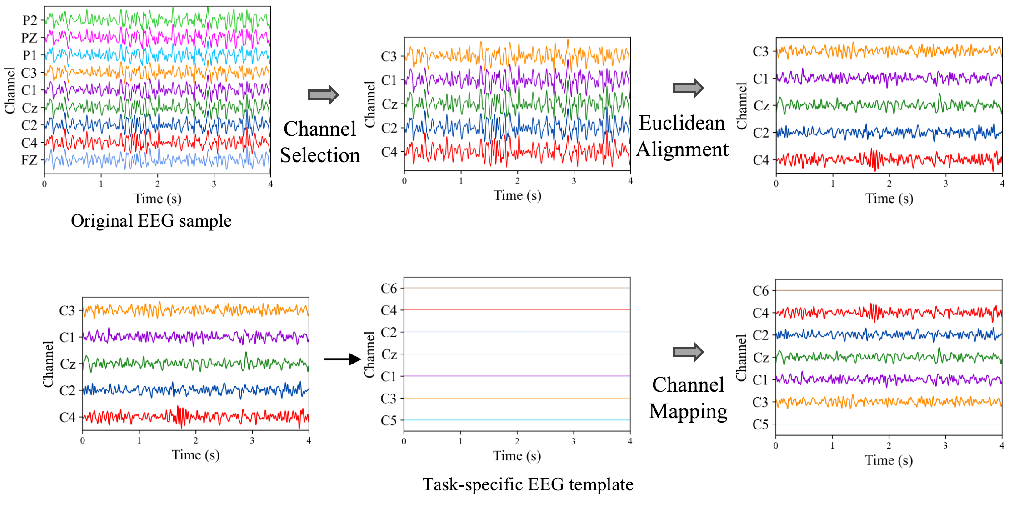}
    \caption{Spatial alignment through channel selection, Euclidean alignment, and channel mapping.}
    \label{fig:sa}
\end{figure*}

The proposed AFPM framework consists of two core modules: Spatial Alignment (SA) and Frame Patch Encoding (FPE). SA selects task-relevant EEG channels based on neurophysiological priors (e.g., brain regions associated with task execution), performs Euclidean alignment to reduce cross-domain variability, and remaps selected channels into a unified spatial layout, as illustrated in Fig.~\ref{fig:sa}. FPE segments temporally synchronized EEG data into patches that preserve multi-channel interactions. A Transformer encoder then models global temporal dependencies for classification, as shown in Fig.~\ref{fig:fpe}. AFPM mitigates the need for dataset-specific calibration by integrating neurophysiology-guided alignment with structured patch encoding.

\subsection{Spatial Alignment (SA)}
SA utilizes EEG paradigm-specific brain region priors and addresses EEG data heterogeneity across different datasets, originated from different device configurations. It consists of three stages: channel selection, Euclidean alignment, and channel mapping, standardizing EEG inputs across datasets with varying sensor configurations and durations.

\subsubsection{Channel Selection} 
Task-relevant neural activity is often localized to specific EEG channels. For example, MI primarily activates the primary motor cortex. Selecting these channels enhances decoding efficiency by emphasizing electrophysiologically meaningful signals.

Let $X\in \mathbb{R}^{N\times T}$ be an EEG sample with $N$ channels and $T$ time domain samples, $\mathcal{S}=\{s_1,s_2,...,s_N\}$ the original channel set of $X$, $\mathcal{T}=\{t_1,t_2,...,t_M\}$ the target channel set for task specific EEG decoding. EEG sample after channel selection  $X'$ is derived through:
\begin{align}
    X'=X[\mathcal{C},:],
    \label{alg:cs}
\end{align}
where $\mathcal{C}=\mathcal{S}\bigcap \mathcal{T}$ is the selected channel set. Note that the original channel sets  differ across datasets, and the target channel sets vary across EEG tasks.

\subsubsection{Euclidean Alignment} 
Given the non-stationarity of EEG signals, after channel selection, Euclidean space alignment is then performed to mitigate inter-domain discrepancy~\cite{He2019}. For $D$ channel-selected EEG samples in a particular domain $\{X'_n\}_{n=1}^D$, i.e. EEG samples collected at the same time period, with the same EEG cap, from the same subject, the Euclidean arithmetic mean $\bar{R}$ of their spatial covariance matrices is:
\begin{align}
\bar{R}=\frac{1}{D} \sum_{n=1}^{D} X'_n\left(X'_n\right)^{\top}.\label{alg:ea1}
\end{align}
Then, the aligned EEG sample $\tilde{X'_n}$ is obtained by:
\begin{align}
\tilde{X}'_n=\bar{R}^{-1 / 2} X'_n, \quad n=1,...,D
\label{alg:ea2}
\end{align}
Euclidean space alignment enforces covariance standardization such that $\frac{1}{D} \sum_{n=1}^D \tilde{X}_n^{\prime \top} \approx \mathbf{I}_M
$, where $\mathbf{I}_M$ denotes $M$ dimensional identity matrix.  As shown in Fig.~\ref{fig:sa}, the original EEG signals across different channels exhibit similar waveforms and are potentially subject to common noise interference. After Euclidean alignment, the waveforms demonstrate more distinct variations across channels, with the signals exhibiting smoother/more stable waveforms.

\subsubsection{Channel Mapping} 
EEG channels exhibit task-specific sensitivities; thus, standardizing channel positions facilitates effective model training.

For a given EEG classification task, channel mapping initiates by constructing a task-specific EEG template matrix $X^{tem}\in \mathbb{R}^{M\times T'}$ with $M$ channels and $T'$ time domain samples. $X^{tem}$ is intialized with 0. Given an aligned EEG sample $\tilde{X'}$, channel mapping is performed by:
\begin{align}
    X^{tem}[\mathcal{C},:T ]\leftarrow\tilde{X'},
    \label{alg:cs}
\end{align}
where $\mathcal{C}=\mathcal{S}\bigcap \mathcal{T}$ is the intersection set of task-specific target channel set $\mathcal{S}$ and EEG sample's original channel set $\mathcal{T}$, and usually  $T'\geq T$. This mapping operation enables neural networks to receive structurally standardized inputs with consistent channel layout, facilitating effective extraction of task-relevant information from cross-dataset EEG representations with varying channels.

In summary, SA removes irrelevant channels, enhances signal quality, and harmonizes EEG representations through whitening, aligning domain-specific signals to a shared space while preserving neurophysiological structure.

\subsection{Frame Patch Encoding (FPE)}
\begin{figure}
    \centering
    \includegraphics[width=0.7\linewidth]{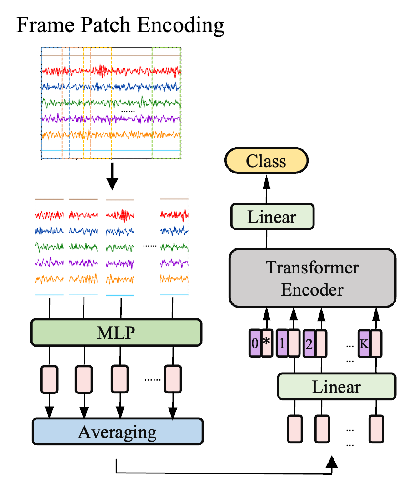}
    \caption{Frame Patch Encoding and Transformer encoder.}
    \label{fig:fpe}
\end{figure}

To enhance the model's efficiency in extracting task-relevant information from channel activities in EEG, FPE processes the multi-channel EEG signals over temporal windows, effectively extracting task-related information characterized by multi-channel representations. As shown in Fig.~\ref{fig:fpe}.

\subsubsection{Frame Patch Encoding}
EEG signals represent spatiotemporal patterns of brain activity. A single EEG channel is insufficient to capture the full scope of neural dynamics associated with cognitive or motor tasks. Therefore, rather than dividing individual channels or subsets into patches, it is more effective to construct patches using the full set of multichannel EEG signals. This approach allows the model to exploit inter-channel dependencies and extract richer task-relevant features.

After channel mapping, FPE maps $X^{tem}$ to a set of embeddings $\mathcal{E} = \{\bm{e}_g|\bm{e}_g\in \mathbb{R}^{L},g \in \{1,2,...,G\}\}$, where $L$ is the dimension of the embedding. Each embedding $\bm{e}_g$ is mapped from all the EEG channel samples over a temporal window from time point $(g-1)d$ to $(g-1)d+m$ through multilayer perceptron $\text{MLP}$:
\begin{align}
    \bm{e}_g=\text{MLP}(X^{tem}[:,(g-1)d:(g-1)d+m]),
\end{align}
where $d$ is the frame stride and $m$ the frame patch window length. $G=\lceil T'/d \rceil +1$ is the number of total embeddings. Zero-padding is applied at matrix boundaries to prevent index overflow, with appropriate reshaping operations ensuring dimensional compatibility for MLP processing.

Given the non-stationarity of EEG signals, FPE further averages $P$ adjacent embeddings to obtain  $\tilde{\bm{e}}_k$ by:
\begin{align}
\tilde{\bm{e}}_k = \frac{1}{P}\sum_{i=1}^P \bm{e}_{(k-1)h+i},
\end{align}
where $h$ is the shifting step between averaging windows, and $P$ the window length parameter. The averaged embedding set $\tilde{\mathcal{E}} = \{\tilde{\bm{e}}_k|\tilde{\bm{e}}_k\in \mathbb{R}^{L},k \in \{1,2,...,K\}\}$, where $K=\lfloor (G-P)/h\rfloor+1$ is the total number of averaged embeddings.

\subsubsection{Class Token and Position Embedding}
To help the model learn the temporal and global information of EEG signals, the averaged embedding set $\tilde{\mathcal{E}}$, temporal position embedding $E_{pos}\in \mathbb{R}^{(K+1)\times L'}$ and class token $\bm{e}_{cls}\in \mathbb{R}^{L'}$ are integrated to form the embedding $E$:
\begin{align}
    E=[\bm{e}_{cls};\tilde{\bm{e}}_1E_0;\tilde{\bm{e}}_2E_0;...;\tilde{\bm{e}}_KE_0]+E_{pos},
\end{align}
where $E_0 \in \mathbb{R}^{L\times L'}$ projects embeddings to transformer token dimension $L'$, $E_{pos}$ encodes temporal ordering and $\bm{e}_{cls}$ serves as a learnable classification token.

\subsubsection{Transformer Encoder}
The embedding $E\in \mathbb{R}^{(K+1)\times L'}$ is fed into Transformer encoder to obtain the contextualized output embedding $\tilde{E}\in \mathbb{R}^{(K+1)\times L'}$: 
\begin{align}
    \tilde{E}=\text{Transformer}(E),
\end{align}
where the Transformer consists of alternating layers of identical multiheaded selfattention and MLP blocks, LayerNorm is applied before every block, and residual connections after every block. $\tilde{E}_{cls}$ is then projected through a learnable linear transformation $W \in \mathbb{R}^{ L' \times c} $ to obtain the logits for classification:
\begin{align}
    \text{logits}=W^{\top}\cdot\tilde{E}_{cls}.
\end{align}

\section{Experiments} \label{sect:es}

\subsection{Datasets and Preprocessing}

To validate AFPM's cross-dataset decoding performance, we trained and tested the model on EEG datasets within the same or similar paradigm. 

\subsubsection{Motor Imagery (MI) and Motor Execution (ME) Datasets} 
MI and ME BCIs are based on event-related desynchronization and synchronisation phenomena in sensorimotor rhythms. During kinesthetic imagination or execution of limb movements (e.g., left/right hand), contralateral sensorimotor cortex activity exhibits event-related desynchronization (amplitude attenuation). In contrast, ipsilateral activity shows event-related synchronisation (amplitude enhancement). These distinct neural patterns enable the classification of imagined motions. 

For MI and ME tasks, PhysionetMI~\cite{Costa2003}, Dreyer2023~\cite{Dreyer2023}, Stieger2021~\cite{Stieger2021}, Cho2017~\cite{Cho2017}, Lee2019-MI~\cite{Lee2019}, Schirrmeister2017~\cite{Schirrmeister2017}, M3CV~\cite{Huang2022}, and BNCI2014-004~\cite{Leeb2007} were used together as the training set, and BNCI2014-001~\cite{Tangermann2012}, Weibo2014~\cite{Yi2014}, and Zhou2016~\cite{Zhou2016} as test sets. MI/ME datasets are usually class-balanced. Key characteristics of MI/ME datasets, including the number of channels, subjects, sessions, trials, and trial length, are summarized in Table~\ref{tab:MI}. 

\begin{table*}[htbp]
\small
  \centering
  \caption{Summary of MI datasets}\label{tab:MI}
    \begin{tabular}{cccccccc}
    \toprule
    Usage & Datasets & \# Channels & \# Subjects & \# Sessions & \# Trials  & Trial length (s) & Classification task \\
    \midrule
    \multirow{8}[2]{*}{Training} & PhysionetMI & 64    & 109   & 1     & 9845  & 3 & left/ right hand \\
          & Dreyer2023 & 27    & 87    & 1     & 20792 & 5& left/ right hand \\
          & Stieger2021 & 64    & 62    & 7 or 11 & 77471 & 3 & left/ right hand \\
          & Cho2017 & 64    & 52    & 1     & 7680  & 3 & left/ right hand\\
          & Lee2019-MI & 62    & 54    & 2     & 10900 & 4 & left/ right hand\\
          & Schirrmeister2017 & 128   & 14    & 1     & 6742  & 4 & left/ right hand\\
          & M3CV  & 64    & 95    & 1     & 15228 & 4 & left/ right hand\\
          & BNCI2014-004 & 3     & 9     & 5     & 6520  & 4.5 & left/ right hand\\
    \midrule
    \multirow{3}[2]{*}{Test} & BNCI2014-001 & 22    & 9     & 2     & 2592  & 4 & left/ right hand\\
          & Weibo2014 & 60    & 10    & 1     & 1580  & 4 & left/ right hand\\
          & Zhou2016 & 14    & 4     & 3     & 1199  & 5 & left/ right hand\\
    \bottomrule
    \end{tabular}%
  \label{tab:addlabel}%
\end{table*}%

\subsubsection{Event-related potentials (ERPs) Datasets}
ERPs are voltage fluctuations in EEG signals time-locked to specific sensory, cognitive, or motor events. ERPs are isolated event-specific neurophysiological responses from background neural activity. They reflect distinct components (e.g., P300) with characteristic latencies and amplitudes. 

For ERP paradigm, BI2012, BI2013a, BI2014a, BI2014b, BI2015a, and BI2015b~\cite{Vaineau2019,Korczowski2019,Korczowski2019b} were used as training set, with BNCI2015-003~\cite{Guger2009} and EPFLP300~\cite{Hoffmann2008} as test sets. Key characteristics of ERP datasets, including the number of channels, subjects, sessions, trials, and trial length, are summarised in Table~\ref{tab:ERP}. ERP datasets are class-imbalanced, and the target to non-target ratio of ERP datasets is around 1:5.

\begin{table*}[htbp]
\small
  \centering
  \caption{Summary of ERP datasets}\label{tab:ERP}
    \begin{tabular}{cccccccc}
    \toprule
    Usage & Datasets & \# Channels & \# Subjects & \# Sessions & \# Trials  & Trial length (s) & Classification task \\
    \midrule
    \multirow{6}[2]{*}{Training} & BI2012 & 16    & 25    & 2     & 19144 & 1 & target/ non-target\\
          & BI2013a & 16    & 24    & 1 or 8 & 35064 & 1 & target/ non-target\\
          & BI2014a & 16    & 64    & up to 3 & 59734 & 1 & target/ non-target\\
          & BI2014b & 32    & 38    & 3     & 8388  & 1 & target/ non-target\\
          & BI2015a & 32    & 43    & 3     & 71484 & 1 & target/ non-target\\
          & BI2015b & 32    & 44    & 1     & 116098 & 1 & target/ non-target\\
    \midrule
    \multirow{2}[2]{*}{Test} & BNCI2015-003 & 8     & 10    & 1     & 25200 & 0.8 & target/ non-target\\
          & EPFLP300 & 32    & 8     & 4     & 26316 & 1 & target/ non-target\\
    \bottomrule
    \end{tabular}%
  \label{tab:addlabel}%
\end{table*}%

\subsubsection{Preprocessing} 
To mitigate noise while preserving task-relevant information, we applied band-pass filters with 4-30 Hz frequency ranges for MI/ME datasets and 1-30 Hz for ERP datasets. Additionally, all data were resampled to 256 Hz. As the range of EEG value is typically between -0.1 mV to 0.1 mV, we set the unit to 0.1 mV to guarantee the value mainly between -1 to 1. %

\subsection{Experiment Settings}

\subsubsection{Pretraining Setting} 
In pretraining, the AdamW optimizer was employed with the OneCycle learning rate strategy (initial learning rate of 2.5e-4, maximum of 5e-4). The training was conducted for 50 epochs, with a batch size of 512 and 16-bit mixed precision training on 4 Nvidia A800 GPUs. Given the significant class imbalance in the ERP dataset, class-balanced sampling was implemented during model training. Our code will be released soon.

\subsubsection{Model Parameters} 
For MI paradigm, we used target channel set $\mathcal{T}^\text{MI}=\{$FC3, FC1, FCZ, FC2, FC4, C5, C3,C1, CZ, C2, C4, C6, CP3, CP1, CPZ, CP2, CP4$\}$ including 17 channels covering the primary motor cortex of the brain related to MI tasks. For ERP paradigm, we used target channel set $\mathcal{T}^\text{ERP}=\{$ FP1, FP2, F5, F3, FZ, F4, F6, FCZ, T7 ,C3, CZ, C4, T8, CP3, CPZ, CP4, P7, P3, PZ, P4, P8, PO7, PO3, PO4, PO8, O1, OZ, O2$\}$ including 28 channels covering midline parietal lobe and central region of the brain related to ERP tasks. Model hyper-parameters for MI and ERP paradigm are shown in Table~\ref{tab:para}.

\begin{table}[htbp]
  \centering
  \caption{Model hyper-parameters for MI and ERP paradigm. 'depth' is the number of layers of the Transformer encoder. 'head' is the number of heads in Transformer and 'mlp\_dim' the intermediate dimension of each transformer layer. 'dim\_head' is the dimension of each head.}
    \begin{tabular}{ccccccccc}
    \toprule
      & $L$ & $m$ & $P$ & $h$ & depth & heads & dim\_head & dim\_mlp \\
    \midrule
    MI & 20 & 25 & 25 & 5 & 6 & 8 & 64 & 40 \\
    ERP & 20 & 25 & 5 & 2 & 6 & 8 & 10 & 20 \\
    \bottomrule
    \end{tabular}%
  \label{tab:para}%
\end{table}%

\subsubsection{Baselines} 
We compare AFPM with non-foundation and foundation models for a comprehensive evaluation. For non-foundation models, we adopt FFCL~\cite{Li2022a}, SPaRCNet~\cite{Jing2023}, ContraWR~\cite{Yang2023a}, CNNTransformer~\cite{peh2022transformer}, STTransformer~\cite{Song2021}, EEGNet~\cite{Lawhern2018}, DeepCNN~\cite{Schirrmeister2017}, ShallowCNN~\cite{Schirrmeister2017}, EEGConformer~\cite{Song2023}, IFNet~\cite{Wang2023d}, FBCNet~\cite{Mane2021}, Deformer~\cite{Ding2025}. Besides, we use BENDER~\cite{Kostas2021}, BIOT~\cite{Yang2023}, LaBraM~\cite{Jiang2024}, EEGPT~\cite{Wang2024}, and CBraMod~\cite{Wang2025} as the foundation-model baselines.

\subsubsection{Evaluation Setting}
AFPM was evaluated directly on downstream datasets using the pretrained model without fine-tuning. We performed cross-subject three-fold cross-validation for all baseline methods and reported the average performance across all folds. Foundation model baselines were initialised using their publicly released pretrained weights before conducting the same cross-validation procedure.

To ensure fairness, baseline models were trained and calibrated using each dataset's original channel configurations and temporal durations, preserving the native data structure. Unless otherwise specified in the ablation study, Euclidean alignment~\cite{He2019,drwuMITLBCI2022} was applied to all datasets across all experiments. Each experiment was repeated three times, and the mean and standard deviation of the results are reported.

\subsubsection{Metrics} 
We adopted Balanced Accuracy and AUC-PR for class-balanced classification, i.e., MI classification, and AUROC, AUC-PR, and Cohen's Kappa for class-imbalanced classification, i.e. ERP classification. Balanced Accuracy is the average recall in each class. AUROC is the area under the receiver operating characteristic curve. Cohen's Kappa is a measure of agreement between categorical variables. AUC-PR is a performance metric that calculates the area under the precision-recall curve.

\subsection{Results on MI}
We compared the effectiveness of AFPM with 12 non-foundation models and 5 EEG foundation models pretrained on upstream datasets. All the baseline approaches were trained or calibrated on target datasets using cross-subject three-fold cross-validation, and AFPM was not calibrated. The experimental results on three MI datasets are summarized in Table~\ref{tab:res_mi}. Observe that:
\begin{enumerate}
    \item  Our AFPM framework consistently outperforms all 17 baseline methods across three MI datasets without any dataset-specific calibration. AFPM achieved a 4.4\% higher Balanced Accuracy and 6.78\% improvement in AUC-PR over the strongest baseline on Weibo2014 dataset.
    \item  Both non-foundation models and foundation models achieved competent performance in MI classification. EEG Foundation model CBraMod achieved the highest performance on BNCI2014-001 among 17 baseline approaches, while non-foundation model IFNet/Conformer was the best on Weibo2014/Zhou2016. 
    \item  Among the 12 non-foundation models, ShallowCNN yielded optimal results on BNCI2014-001. Within the 5 EEG foundation models, CBraMod attained the highest metrics on BNCI2014-001 and Weibo2014, while LaBraM achieved the best performance on Zhou2016. 
\end{enumerate}

\begin{table*}[htbp]
  \centering
    \footnotesize
\setlength{\tabcolsep}{1mm}
  \caption{Results of different approaches on MI datasets. 'Pre' stands for 'Pretrained'. 'CF' stands for 'Calibration-free'. The largest values in each column are marked in bold. The largest values in each column for a category of approaches are underlined.}
    \begin{tabular}{cccrccrccrcc}
    \toprule
    \multirow{2}[4]{*}{Approach} & \multicolumn{1}{c}{\multirow{2}[4]{*}{Pre}} & \multicolumn{1}{c}{\multirow{2}[4]{*}{CF}} &       & \multicolumn{2}{c}{BNCI2014-001} &       & \multicolumn{2}{c}{Weibo2014} &       & \multicolumn{2}{c}{Zhou2016} \\
\cmidrule{5-6}\cmidrule{8-9}\cmidrule{11-12}          &       &       &       & Balanced Accuracy & AUC-PR &       & Balanced Accuracy & AUC-PR &       & Balanced Accuracy & AUC-PR \\
    \midrule
    FFCL  &     \ding{55}  &   \ding{55}     &       & 0.7095 $\pm$ 0.0095 & 0.7750 $\pm$ 0.0048 &       & 0.6910 $\pm$ 0.0087 & 0.7410 $\pm$ 0.0063 &       & 0.7972 $\pm$ 0.0038 & 0.8860 $\pm$ 0.0079 \\
    SPaRCNet &    \ding{55}  &   \ding{55}      &       & 0.6623 $\pm$ 0.0070 & 0.7262 $\pm$ 0.0046 &       & 0.6266 $\pm$ 0.0132 & 0.6895 $\pm$ 0.0066 &       & 0.8054 $\pm$ 0.0099 & 0.8969 $\pm$ 0.0092 \\
    ContraWR &     \ding{55}  &   \ding{55}      &       & 0.7316 $\pm$ 0.0116 & 0.8022 $\pm$ 0.0078 &       & 0.7078 $\pm$ 0.0177 & 0.7678 $\pm$ 0.0125 &       & 0.8583 $\pm$ 0.0073 & 0.9449 $\pm$ 0.0033 \\
    CNNTransformer &     \ding{55}  &   \ding{55}     &       & 0.6883 $\pm$ 0.0107 & 0.7675 $\pm$ 0.0085 &       & 0.6581 $\pm$ 0.0085 & 0.7403 $\pm$ 0.0156 &       & 0.8429 $\pm$ 0.0029 & 0.9319 $\pm$ 0.0021 \\
    STTransformer &    \ding{55}  &   \ding{55}      &       & 0.6335 $\pm$ 0.0103 & 0.6880 $\pm$ 0.0061 &       & 0.5339 $\pm$ 0.0033 & 0.5580 $\pm$ 0.0106 &       & 0.7845 $\pm$ 0.0223 & 0.8836 $\pm$ 0.0203 \\
    EEGNet &     \ding{55}  &   \ding{55}     &       & 0.7494 $\pm$ 0.0083 & 0.8253 $\pm$ 0.0037 &       & 0.6772 $\pm$ 0.0372 & 0.7382 $\pm$ 0.0590 &       & 0.8549 $\pm$ 0.0088 & 0.9395 $\pm$ 0.0086 \\
    DeepCNN &   \ding{55}  &   \ding{55}     &       & 0.7128 $\pm$ 0.0069 & 0.7858 $\pm$ 0.0018 &       & 0.5984 $\pm$ 0.0310 & 0.6205 $\pm$ 0.0301 &       & 0.8562 $\pm$ 0.0065 & 0.9345 $\pm$ 0.0039 \\
    ShallowCNN &    \ding{55}  &   \ding{55}     &       & \underline{0.7761} $\pm$ 0.0063 & \underline{0.8593} $\pm$ 0.0052 &       & 0.6929 $\pm$ 0.0035 & 0.7662 $\pm$ 0.0007 &       & 0.8642 $\pm$ 0.0142 & 0.9492 $\pm$ 0.0017 \\
    Conformer &     \ding{55}  &   \ding{55}      &       & 0.7460 $\pm$ 0.0043 & 0.8276 $\pm$ 0.0023 &       & 0.7093 $\pm$ 0.0107 & 0.7749 $\pm$ 0.0161 &       & \underline{0.8648} $\pm$ 0.0082 & \underline{0.9514} $\pm$ 0.0046 \\
    IFNet &   \ding{55}  &   \ding{55}   &       & 0.7468 $\pm$ 0.0071 & 0.8231 $\pm$ 0.0108 &       & \underline{0.7180} $\pm$ 0.0107 & \underline{0.7833} $\pm$ 0.0159 &       & {0.8607} $\pm$ 0.0031 & {0.9514} $\pm$ 0.0038 \\
    FBCNet &  \ding{55}  &   \ding{55}    &       & 0.6929 $\pm$ 0.0016 & 0.7570 $\pm$ 0.0013 &       & 0.6562 $\pm$ 0.0138 & 0.7036 $\pm$ 0.0181 &       & 0.8400 $\pm$ 0.0074 & 0.9229 $\pm$ 0.0055 \\
    Deformer &   \ding{55}  &   \ding{55}   &       & {0.7685} $\pm$ 0.0051 & {0.8418} $\pm$ 0.0081 &       & 0.6407 $\pm$ 0.0119 & 0.6892 $\pm$ 0.0204 &       & 0.8501 $\pm$ 0.0181 & 0.9487 $\pm$ 0.0043 \\
    \midrule
    BENDER &  \checkmark     &  \ding{55}     &       & 0.7140 $\pm$ 0.0086 & 0.7732 $\pm$ 0.0095 &       & 0.5091 $\pm$ 0.0019 & 0.5144 $\pm$ 0.0076 &       & 0.6583 $\pm$ 0.0137 & 0.7232 $\pm$ 0.0068 \\
    BIOT & \checkmark     &  \ding{55}     &       & 0.6592 $\pm$ 0.0090 & 0.7089 $\pm$ 0.0154 &       & 0.5499 $\pm$ 0.0344 & 0.5692 $\pm$ 0.0336 &       & 0.6397 $\pm$ 0.0860 & 0.7093 $\pm$ 0.1300 \\
    LaBraM &  \checkmark     &  \ding{55}    &       & 0.7692 $\pm$ 0.0034 & 0.8514 $\pm$ 0.0030 &       & 0.6822 $\pm$ 0.0101 & 0.7348 $\pm$ 0.0143 &       & \underline{0.8407} $\pm$ 0.0032 & \underline{0.9193} $\pm$ 0.0006 \\
    EEGPT &  \checkmark     &  \ding{55}    &       & 0.5931 $\pm$ 0.0345 & 0.6303 $\pm$ 0.0511 &       & 0.5026 $\pm$ 0.0172 & 0.4962 $\pm$ 0.0135 &       & 0.6353 $\pm$ 0.0915 & 0.7091 $\pm$ 0.1163 \\
    CBraMod &   \checkmark     &  \ding{55}    &       & \underline{0.7870} $\pm$ 0.0020 &\underline{0.8622} $\pm$ 0.0028 &       & \underline{0.6850} $\pm$ 0.0079 & \underline{0.7432} $\pm$ 0.0089 &       & 0.8074 $\pm$ 0.0055 & 0.8864 $\pm$ 0.0025 \\
    \midrule
    AFPM(ours) &   \checkmark    &   \checkmark     &       & \textbf{0.7958} $\pm$ 0.0091 &\textbf{ 0.8678} $\pm$ 0.0075 &       &\textbf{0.7620} $\pm$ 0.0072 & \textbf{0.8511} $\pm$ 0.0056 &       & \textbf{0.8823} $\pm$ 0.0103 & \textbf{0.9663} $\pm$ 0.0050 \\
    \bottomrule
    \end{tabular}%
  \label{tab:res_mi}%
\end{table*}%

\subsection{Results on ERP}
We compared the performance of AFPM with 12 non-foundation models and 5 EEG foundation models pretrained on upstream datasets. All the baseline approaches were trained or calibrated on target datasets using cross-subject three-fold cross-validation, and AFPM was not calibrated. The experimental results on two ERP datasets are in Table~\ref{tab:res_erp}. Observe that:

\begin{enumerate}
    \item Our AFPM approach, without requiring downstream dataset fine-tuning, surpassed all 17 baselines. Specifically, on EPFLP300, AFPM achieved absolute improvements of 2.98\% AUROC, 6.66\% AUC-PR, and 3.78\% Cohen's Kappa over the highest baseline values.
    \item Among the 17 baseline approaches, non-foundation models exhibited marginally superior performance over foundation models on ERP datasets. Specifically, non-foundation models achieved higher AUROC and AUC-PR values across all 17 baselines for EPFLP300 and BNCI2015-003 datasets, except EEGPT, attaining higher Cohen's Kappa on the EPFLP300 dataset.
    \item   Within the 12 non-foundation baseline models, EEGNet was the most effective overall, achieving better performance in both Cohen's Kappa on EPFLP300 and all metrics on BNCI2015-003. Among the five foundation model baselines, EEGPT demonstrated the best overall performance and consistently outperformed others across multiple metrics in both datasets, except that CBraMod attained marginally higher AUC-PR on BNCI2015-003. 
\end{enumerate}

\begin{table*}[htbp]
  \centering
      \footnotesize
\setlength{\tabcolsep}{1.2mm}
  \caption{Results of different approaches on ERP datasets. 'Pre' stands for 'Pretrained'. 'CF' stands for 'Calibration-free'. The largest values in each column are marked in bold. The largest values in each column for a category of approaches are underlined.}
    \begin{tabular}{ccccccccccc}
    \toprule
    \multirow{2}[4]{*}{Approach} & \multirow{2}[4]{*}{Pre} & \multirow{2}[4]{*}{CF} &       & \multicolumn{3}{c}{EPFLP300} &       & \multicolumn{3}{c}{BNCI2015-003} \\
\cmidrule{5-7}\cmidrule{9-11}          &       &       &       & AUROC & AUC-PR & Coken's Kappa &       & AUROC & AUC-PR & Coken's Kappa \\
    \midrule
    FFCL  &      \ding{55}  &   \ding{55}      &       & 0.6020 $\pm$ 0.0046 & 0.2361 $\pm$ 0.0033 & 0.0947 $\pm$ 0.0084 &       & 0.6103 $\pm$ 0.0060 & 0.1912 $\pm$ 0.0040 & 0.0796 $\pm$ 0.0070 \\
    SPaRCNet &     \ding{55}  &   \ding{55}      &       & 0.6156 $\pm$ 0.0056 & 0.2445 $\pm$ 0.0033 & 0.1069 $\pm$ 0.0040 &       & 0.6073 $\pm$ 0.0047 & 0.2050 $\pm$ 0.0078 & 0.0997 $\pm$ 0.0145 \\
    ContraWR &     \ding{55}  &   \ding{55}      &       & 0.5693 $\pm$ 0.0019 & 0.2073 $\pm$ 0.0027 & 0.0601 $\pm$ 0.0028 &       & 0.5082 $\pm$ 0.0084 & 0.1351 $\pm$ 0.0016 & 0.0038 $\pm$ 0.0047 \\
    CNNTransformer &      \ding{55}  &   \ding{55}       &       & 0.5644 $\pm$ 0.0057 & 0.2072 $\pm$ 0.0007 & 0.0641 $\pm$ 0.0054 &       & 0.5022 $\pm$ 0.0093 & 0.1363 $\pm$ 0.0013 & 0.0041 $\pm$ 0.0033 \\
    STTransformer &    \ding{55}  &   \ding{55}      &       & 0.5938 $\pm$ 0.0073 & 0.2283 $\pm$ 0.0054 & 0.0904 $\pm$ 0.0019 &       & 0.5934 $\pm$ 0.0167 & 0.1874 $\pm$ 0.0105 & 0.0625 $\pm$ 0.0062 \\
    EEGNet &    \ding{55}  &   \ding{55}    &       & 0.6337 $\pm$ 0.0064 & 0.2636 $\pm$ 0.0066 & \underline{0.1399} $\pm$ 0.0068 &       & \underline{0.6660} $\pm$ 0.0039 & \underline{0.2482} $\pm$ 0.0038 & \underline{0.1201} $\pm$ 0.0044 \\
    DeepCNN &    \ding{55}  &   \ding{55}     &       & \underline{0.6398} $\pm$ 0.0035 & 0.2681 $\pm$ 0.0024 & 0.1390 $\pm$ 0.0045 &       & 0.6283 $\pm$ 0.0026 & 0.2060 $\pm$ 0.0024 & 0.1067 $\pm$ 0.0056 \\
    ShallowCNN &    \ding{55}  &   \ding{55}      &       & 0.6236 $\pm$ 0.0065 & 0.2518 $\pm$ 0.0050 & 0.1304 $\pm$ 0.0096 &       & 0.6339 $\pm$ 0.0012 & 0.2132 $\pm$ 0.0012 & 0.1038 $\pm$ 0.0041 \\
    Conformer &     \ding{55}  &   \ding{55}     &       & 0.6243 $\pm$ 0.0059 & 0.2455 $\pm$ 0.0061 & 0.1175 $\pm$ 0.0112 &       & 0.5861 $\pm$ 0.0032 & 0.1801 $\pm$ 0.0040 & 0.0736 $\pm$ 0.0096 \\
    IFNet &    \ding{55}  &   \ding{55}     &       & 0.6095 $\pm$ 0.0057 & 0.2413 $\pm$ 0.0055 & 0.1169 $\pm$ 0.0108 &       & 0.5602 $\pm$ 0.0071 & 0.1611 $\pm$ 0.0042 & 0.0483 $\pm$ 0.0014 \\
    Deformer &    \ding{55}  &   \ding{55}    &       & 0.6388 $\pm$ 0.0047 & \underline{0.2688} $\pm$ 0.0069 & 0.1339 $\pm$ 0.0093 &       & 0.6202 $\pm$ 0.0028 & 0.2134 $\pm$ 0.0016 & 0.1092 $\pm$ 0.0030 \\
      \midrule
    BENDER &   \checkmark     &  \ding{55}      &       & 0.6123 $\pm$ 0.0048 & 0.2480 $\pm$ 0.0053 & 0.1027 $\pm$ 0.0065 &       & 0.6219 $\pm$ 0.0090 & 0.2137 $\pm$ 0.0040 & 0.0847 $\pm$ 0.0030 \\
    BIOT &    \checkmark     &  \ding{55}     &       & 0.5279 $\pm$ 0.0031 & 0.1796 $\pm$ 0.0016 & 0.0247 $\pm$ 0.0037 &       & 0.5051 $\pm$ 0.0096 & 0.1341 $\pm$ 0.0048 & -0.0049 $\pm$ 0.0063 \\
    LaBraM &   \checkmark     &  \ding{55}      &       & 0.6151 $\pm$ 0.0011 & 0.2455 $\pm$ 0.0025 & 0.0876 $\pm$ 0.0026 &       & 0.6364 $\pm$ 0.0045 & 0.2244 $\pm$ 0.0038 & 0.1026 $\pm$ 0.0044 \\
    EEGPT &   \checkmark     &  \ding{55}     &       & \underline{0.6338} $\pm$ 0.0062 & \underline{0.2640} $\pm$ 0.0082 & \underline{0.1425} $\pm$ 0.0104 &       & \underline{0.6371} $\pm$ 0.0026 & 0.2320 $\pm$ 0.0037 & \underline{0.1331} $\pm$ 0.0023 \\
    CBraMod &     \checkmark     &  \ding{55}     &       & 0.6148 $\pm$ 0.0018 & 0.2459 $\pm$ 0.0002 & 0.1143 $\pm$ 0.0007 &       & 0.6306 $\pm$ 0.0045 & \underline{0.2360} $\pm$ 0.0039 & 0.1307 $\pm$ 0.0074 \\
    \midrule
    AFPM(ours) &    \checkmark    &   \checkmark    &       & \textbf{0.6696} $\pm$ 0.0019 & \textbf{0.3306} $\pm$ 0.0031 & \textbf{0.1803} $\pm$ 0.0065 &       & \textbf{0.6760} $\pm$ 0.0080 & \textbf{0.2452} $\pm$ 0.0031 & \textbf{0.1620} $\pm$ 0.0082 \\
    \bottomrule
    \end{tabular}%
  \label{tab:res_erp}%
\end{table*}%

\subsection{Ablation Study}
\begin{figure*}
  \centering
  {\includegraphics[width=1\linewidth,clip]{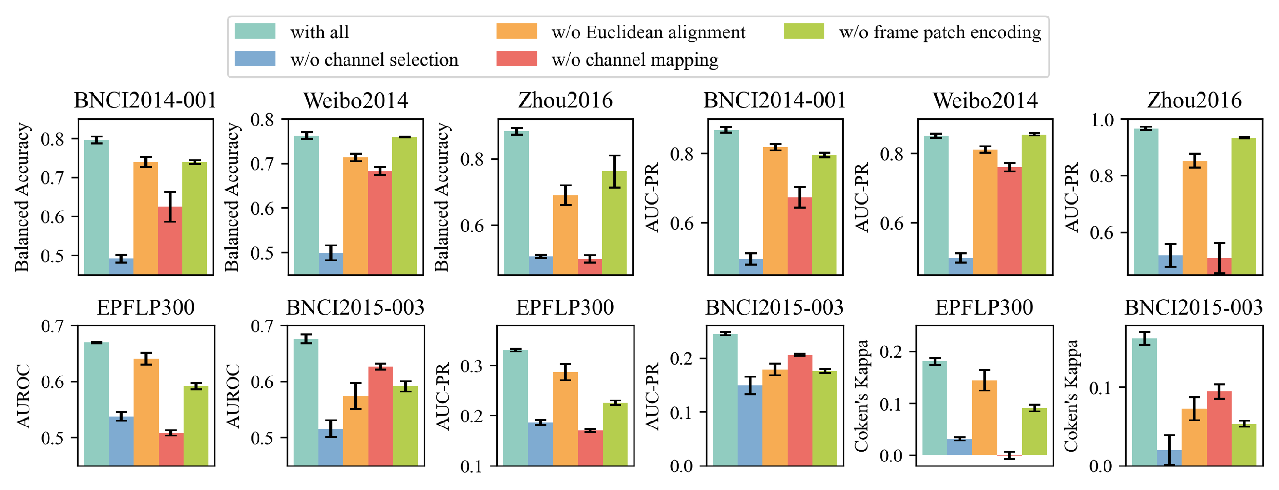}}
  \caption{Results of ablation study on three MI datasets: BNCI2014-001, Weibo2014, Zhou2016, and two ERP datasets: EPFLP300, BNCI2015-003.}
  \label{fig:abla}
\end{figure*}

We conducted ablation experiments under five different configurations. The results on three MI datasets and two ERP datasets are shown in Fig.~\ref{fig:abla}. The "with all" configuration indicates full implementation of all components in AFPM. Four ablation scenarios were constructed by systematically modifying individual modules. In the "w/o channel selection" scenario, all channels of all training datasets are adopted. In the "w/o Euclidean alignment" scenario, all datasets are not aligned in Euclidean space. In the "w/o channel mapping" scenario, each dataset's original channel distribution order is preserved when the EEG data is fed into the network. In the "w/o frame patch encoding" scenario, the signal of each channel of the EEG data for a short period is segmented into a patch. Observe that:
\begin{enumerate}
    \item The channel selection has the most significant impact on MI classification performance in AFPM. Under the "w/o channel selection" scenario, both evaluation metrics on all three datasets show substantial degradation, demonstrating that channel selection effectively enhances EEG signal-to-noise ratio and improves model performance. 
    \item The channel mapping exhibits the second-largest influence on MI classification results, indicating that maintaining structural consistency in EEG input topology plays a critical role in model learning, where standardized electrode arrangements significantly enhance learning efficacy. 
    \item Performance degradation under the "w/o Euclidean alignment" scenario confirms that inherent non-stationarity in EEG data negatively impacts model learning. 
    \item The frame patch encoding also markedly affects AFPM performance, particularly on BNCI2014-001 and Zhou2016 datasets, suggesting that joint encoding of task-relevant EEG channels substantially improves learning efficiency. 
\end{enumerate}

Ablation results in Fig.~\ref{fig:abla} on the two ERP datasets further demonstrate the necessity of these four modules.

\subsection{Model Parameter Analysis}
\begin{figure*}
  \centering
  {\includegraphics[width=1\linewidth,clip]{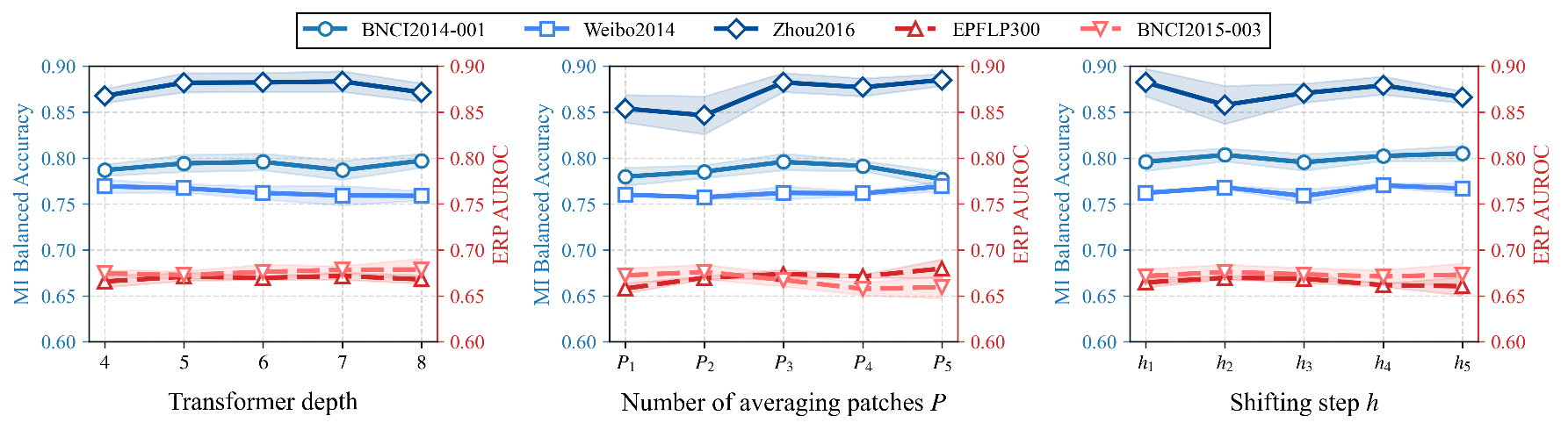}}
  \caption{Influence of model hyper-parameters, i.e., Transformer depth, the number of averaging patches $P$, and shifting step $h$, on model performance on three MI datasets: BNCI2014-001, Weibo2014, Zhou2016, and two ERP datasets: EPFLP300, BNCI2015-003.}
  \label{fig:hyper}
\end{figure*}

The performance of models on the five datasets under different Transformer depth, number of averaging patches ${P}$, and shifting step ${h}$ is shown in Fig.~\ref{fig:hyper}. We chose the Number of averaging patches $P^\text{MI}=\{5,15,25,35,45\}$, shifting steps $h^\text{MI}=\{5,10,15,20,25\}$ for the MI paradigm and $P^\text{ERP}=\{3,5,7,10,15\}$, $h^\text{ERP} = \{1, 2, 3, 4, 5 \} $ for the ERP paradigm.

Model performance remained consistent across five datasets under varying Transformer depths, numbers of averaging patches ${P}$, and shifting steps ${h}$, demonstrating the robustness and stability of our method.

\subsection{Effectiveness of Downstream Fine-tuning}
\begin{figure*}
  \centering
  {\includegraphics[width=0.98\linewidth,clip]{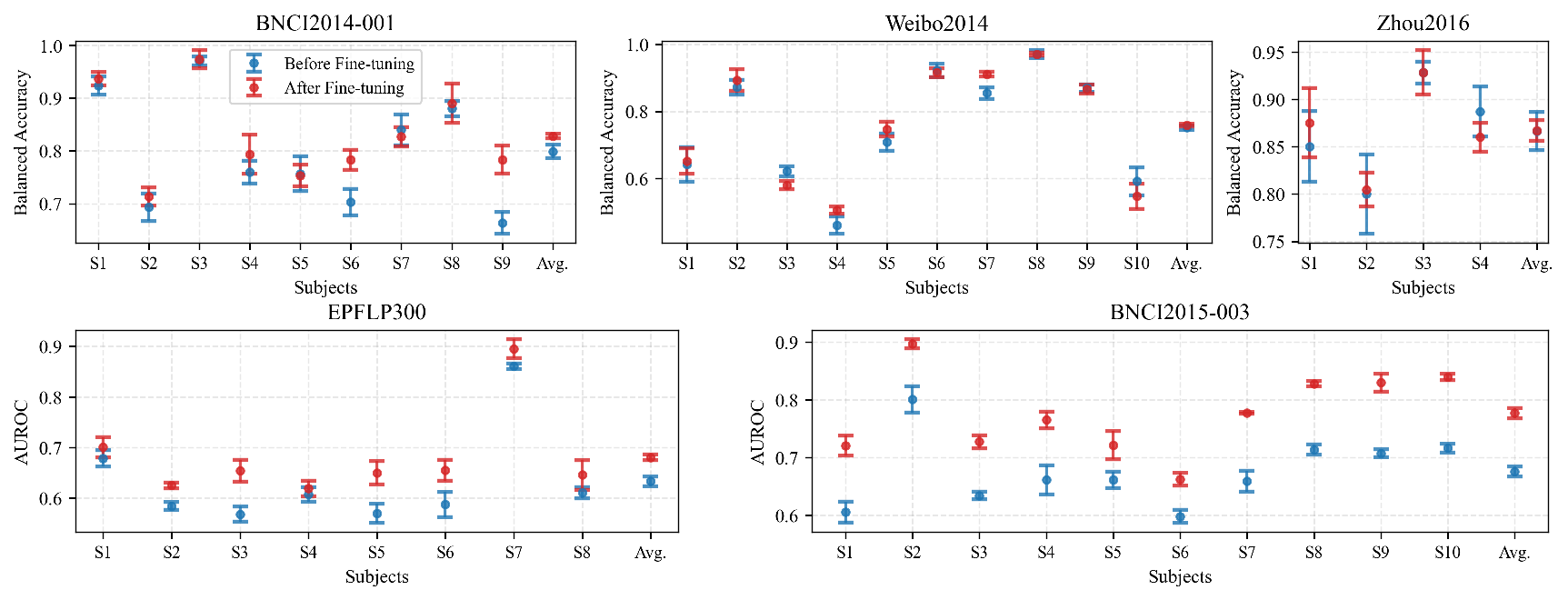}}
  \caption{Subject-wise performance of AFPM before and after subject-specific fine-tuning on three MI datasets: BNCI2014-001, Weibo2014, Zhou2016, and two ERP datasets: EPFLP300, BNCI2015-003.}
  \label{fig:ft}
\end{figure*}

 We performed fine-tuning using 30\% of each subject's data and evaluated model performance on the remaining 70\% before and after fine-tuning. Results across five datasets are in Fig.~\ref{fig:ft}. 
 
It is observed that subject-specific fine-tuning can improve model performance in most cases. Although AFPM has shown good generalisation ability, because the distribution of EEG data varies greatly from subject to subject, subject-specific fine-tuning still significantly affects the EEG decoding performance of the target subject.

\subsection{Influence of Training Data}
\begin{figure}
  \centering
  {\includegraphics[width=1\linewidth,clip]{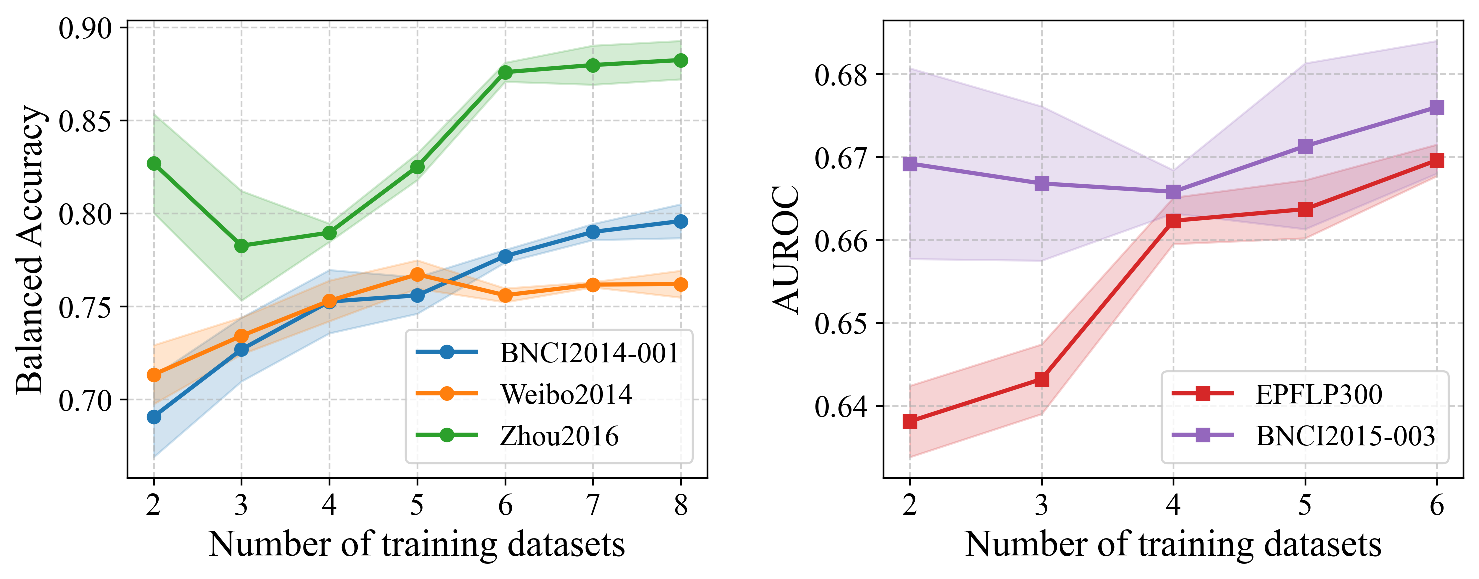}}
  \caption{AFPM Performance on five datasets as the number of training datasets increases.}
  \label{fig:training_data}
\end{figure}

We conducted experiments to investigate the model's performance scaling with the number of training datasets, with results visualised in Fig.~\ref{fig:training_data}. The empirical evidence reveals an overall performance improvement across three MI datasets and two ERP paradigms as more training datasets are incorporated. This demonstrates the critical value of multi-dataset integration in training to enhance EEG decoding performance.

\subsection{Discussion}
Taking the BNCI2014-001 dataset as an example, we compare the parameter counts of various approaches. The results are shown in Table~\ref{tab:para}. Observe that:
\begin{enumerate}
\item Although AFPM has only 269.28k parameters, which is much smaller than those of foundation models (BENDER, BIOT, LaBraM, EEGPT, CBraMod), AFPM can achieve good results in cross-dataset EEG decoding.
\item In non-foundation models, models with smaller parameter counts generally achieved better performance in EEG decoding. For instance, ShallowCNN, with only 35.92k parameters, demonstrated better results in Table~\ref{tab:res_mi}.
\item The parameters of foundation models are usually more than those of non-foundation models because the foundation model usually needs to be large for learning generic representations. 
\end{enumerate}

\begin{table}[htbp]
  \centering
  \setlength{\tabcolsep}{2mm}
  \caption{Model parameters comparison on BNCI2014-001.}
    \begin{tabular}{llll}
    \toprule
    Model & Parameters & Model & Parameters \\
    \midrule
    FFCL & 3.42 M & IFNet & 9.34 k \\
    SPaRCNet & 1.14 M & FBCNet & 1.6 k \\
    ContraWR & 1.57 M & Deformer & 1.04 M \\
    CNNTransformer & 3.16 M & BENDER & 3.98M \\
    STTransformer & 2.64 M & BIOT & 3.19M \\
    EEGNet & 1.43 k & LaBraM & 5.86 M \\
    DeepCNN & 277.65 k & EEGPT & 25.32 M \\
    ShallowCNN & 35.92 k & CBraMod & 4.96 M \\
    Conformer & 156.32 k & AFPM & 269.28 k \\
    \bottomrule
    \end{tabular}%
  \label{tab:para}%
\end{table}%

\section{Conclusions} \label{sect:CFR}
This paper proposes AFPM, a calibration-free framework for cross-dataset EEG decoding, which significantly advances the practical deployment of BCI systems. By integrating spatial alignment through task-informed channel selection, Euclidean alignment, and channel mapping, AFPM effectively mitigates inter-domain distributional shifts and enhances EEG signal quality. The frame-patch encoding module further enables robust modeling of spatiotemporal neural dynamics, resulting in improved decoding performance across heterogeneous datasets. Extensive experiments on MI and ERP tasks demonstrate that AFPM consistently outperforms seventeen state-of-the-art baselines without requiring target-domain fine-tuning or calibration.

This study evaluates AFPM primarily on single-task EEG data. Future work may explore its extension to multi-task EEG foundation model pretraining.

\bibliographystyle{IEEEtran} \bibliography{daat}

\end{document}